\documentclass[aip,amsmath,amssymb,graphicx]{revtex4-1}

\usepackage{dcolumn}
\usepackage{bm}
\usepackage[utf8]{inputenc}
\usepackage[T1]{fontenc}
\usepackage{mathptmx}

\usepackage{float}

\usepackage{amsmath}
\usepackage{mathtools}

\usepackage{siunitx}
\usepackage{float}
\usepackage{xcolor}

\begin{document}

\title{Dual Modulation STM: simultaneous high-resolution mapping of the differential conductivity and local tunnel barrier height demonstrated on Au(111)}

\author{V.J.S. Oldenkotte}
\email{v.j.s.oldenkotte@utwente.nl}
\affiliation{Physics of Interfaces and Nanomaterials, MESA+ Institute for Nanotechnology, University of Twente, P.O. Box 217, 7500 AE Enschede, The Netherlands}
\affiliation{Industrial Focus Group XUV Optics, MESA+ Institute for Nanotechnology, University of Twente, P.O. Box 217, 7500 AE Enschede, The Netherlands}

\author{F.J. Witmans}
\author{M.H. Siekman}
\author{P.L. de Boeij}
\author{K. Sotthewes}
\author{C. Castenmiller}
\affiliation{Physics of Interfaces and Nanomaterials, MESA+ Institute for Nanotechnology, University of Twente, P.O. Box 217, 7500 AE Enschede, The Netherlands}

\author{M.D. Ackermann}
\author{J.M. Sturm}
\affiliation{Industrial Focus Group XUV Optics, MESA+ Institute for Nanotechnology, University of Twente, P.O. Box 217, 7500 AE Enschede, The Netherlands}

\author{H.J.W. Zandvliet}
\affiliation{Physics of Interfaces and Nanomaterials, MESA+ Institute for Nanotechnology, University of Twente, P.O. Box 217, 7500 AE Enschede, The Netherlands}

\date{\today}

\begin{abstract}
    This article may be downloaded for personal use only. Any other use requires prior permission of the author and AIP Publishing. This article appeared in \textit{Journal of Applied Physics}, 129, 225301 (2021) and may be found at https://doi.org/10.1063/5.0051403

    We present a scanning tunneling microscopy (STM) technique to simultaneously measure the topography, the local tunnel barrier height (d$I$/d$z$) and the differential conductivity (d$I$/d$V$). We modulate the voltage and tip piezo with small sinusoidal signals that exceed the cut-off frequency of the STM electronics and feed the tunneling current into two lock-in amplifiers (LIAs). We derive and follow a set of criteria for the modulation frequencies to avoid any interference between the LIA measurements. To validate the technique, we measure Friedel oscillations and the subtle tunnel barrier difference between the $hcp$ and $fcc$ stacked regions of the Au(111) herringbone reconstruction. Finally, we show that our method is also applicable to open feedback loop measurements by performing grid $I$($V$) spectroscopy.
\end{abstract}

\maketitle 

\section{Introduction}

    Almost half a century ago, the scanning tunneling microscope (STM) was invented by Binnig and Rohrer. \cite{binnig1982surface} The advent of the STM revolutionized the way humanity saw and interacted with the world: the resolution of eyes and dexterity of our hands could now be extended down to the atomic scale. Soon after the discovery of STM, new scientific techniques emerged. Among them was scanning tunneling spectroscopy (STS). \cite{zandvliet2009scanning,binnig2000scanning,feenstra1994scanning}
    
    In principle, STS provides electronic information at the atomic scale. The two best-known STS modes are $I$($V$) and $I$($z$) spectroscopy. In the first mode, the tip position is fixed, the bias voltage is swept and the tunnel current is measured. \cite{dasgupta2017band,kubby1987tunneling,hamers1986surface,wedig2016nanoscale} The differential conductivity (d$I$/d$V$) is then proportional to the local density of states (LDOS). In the $I$($z$) spectroscopy mode, the tip-sample separation ($z$) is ramped at constant sample bias, while the current is recorded. This curve offers insight into the apparent tunnel barrier ($\phi$) through its inverse decay length ($\kappa (V)$), which in turn gives information on surface dipoles and parallel momentum. \cite{feenstra1987tunneling,stroscio1986electronic,lang1988apparent,wang2019real}
    
    Variable $z$ spectroscopy techniques, such as $I$($z$) spectroscopy, offer insight into phenomena such as the intricate influence of the parallel momentum on the tunneling process. As an illustrative example of this we refer to work by Zhang et al. \cite{zhang2015probing} By combining constant $z$ spectroscopy, variable $z$ spectroscopy,  and state-resolved tunneling decay constant measurements, information on the dispersion of the energy bands of several transition metal dichalcogenides were extracted.
    
    Measuring the spatial variation of electronic properties, and its correspondence to topographical features, is a powerful tool for surface characterisation. A common way to do this, is to set out a grid of points for a topography scan at which the STM will stop to take a full $I$($V$) or $I$($z$) curve. Consequently, the measurement time can increase tremendously: from minutes up to hours, depending on the desired resolution in the STS grid. This can lead to serious issues with the stability of the tunnel junction and often requires extra architectural investments to ensure the stability of the lab floor and the STM setup.
    
    Another way is to use a lock-in amplifier (LIA) to modulate the variable that is normally ramped and directly measure the differential of the STS curve. \cite{le2017scanning,ruby2015experimental,becker1985electron} By recording this signal during a topography scan one directly measures the spectroscopic map. \cite{tsvetanova2020nanoscale,kundu2017differential,de2008spatial,becker1985real} Although the scanning speed has to be slowed down to ensure a decent resolution in the spectroscopic map, as the response time of the filter inside the LIA can be several times its time constant ($\tau$),\cite{horowitz1989art} this still offers a heavily reduced scan time compared to grid STS scans. The caveat here is that, while taking full STS curves in a grid gives information over a large range of the ramped variable, STS maps only offer information about the STS curve at a single point. For instance, in the case of a d$I$/d$V$ map, this point would be the bias voltage. So, one has to know beforehand at which bias voltage interesting behavior takes place. 
    
    To fully characterize a surface multiple physical and chemical properties have to be determined. Bearing this in mind, combining STS methods can be of great benefit. One way of capturing d$I$/d$V$ and d$I$/d$z$ simultaneously is to combine the two methods: measure one property with an STS map while measuring the other with a grid of STS curves and numerically deriving a map of the latter property. \cite{castenmiller2018combined} Since this is much faster than strictly using grid STS, thermal drift and tip changes will play a smaller role, which makes interpretation easier while still offering the option to explore a broader range of settings. However, the numerically derived map will have a very limited resolution. To get subnanometre resolution in both maps and allow for highly accurate comparisons between features in the topography and spectroscopic maps, the next logical step is dual modulation STM, where d$I$/d$V$ and d$I$/d$z$ maps are captured simultaneously by modulating both $V$ and $z$ and using two LIAs. Of course, this raises the question: how can we ensure that the two STS measurements will not interfere?
    
    Here, we show that this dual modulation STM method works and, as long as one carefully picks the settings of the LIAs, no interference between the spectroscopic measurements occurs. We derive a set of criteria for proper usage of this method and illustrate its performance for the well-studied Au(111) surface.

\section{Theory}
The tunneling current can be described within the Tersoff-Hamann model, \cite{tersoff1983theory} shown below in Equation \ref{th_eq_tersoff_hamann}, where $\rho(\epsilon,z)$ is the LDOS, $\epsilon$ is the energy of electronic states that participate in the tunneling process with respect to the Fermi energy, $V$ is the applied bias voltage, $z$ is the tip-sample distance, and $\kappa (V)$ is the voltage dependent inverse decay length of the tunnel barrier.

\begin{equation}
\label{th_eq_tersoff_hamann}
\begin{split}
    I(V,z) =& C\int_{0}^{eV} \rho(\epsilon,V,z)d\epsilon, \text{ Where:}\\
    \rho(\epsilon,V,z) \sim& \rho(\epsilon,V,0) \exp(-2\kappa(V) z)
\end{split}
\end{equation}

In this general expression for the tunnel current we have not considered a single electronic state, but rather consider all the electronic states in the energy window [0,eV] that contribute to the tunneling process. For a more detailed analysis where we have considered all the involved electronic states, we refer to Equations \ref{app_eq_rho} and \ref{app_eq_simplify_rho} in Appendix (Section \ref{app_lia}). As a consequence of this, the measured inverse decay length can be regarded as a weighted average of the inverse decay lengths of all states participating in the tunneling process. In some cases, however, this issue can be safely disregarded as only a single state participates in the tunneling process, such as for graphene on MoS\textsubscript{2}. \cite{jiao2021}

To measure the d$I$/d$V$ and d$I$/d$z$, we modulate $V$ and $z$ with two periodic functions: $V = V_0 + \hat{v}\sin(\omega_V t + \phi_V)$ and $z = z_0 + \hat{z}\sin(\omega_z t + \phi_z)$, respectively. In the LIA, the tunneling current is multiplied by a reference signal: $\sin(\omega_V t + \phi_V)$ for the d$I$/d$V$ and $\sin(\omega_z t + \phi_z)$ for the d$I$/d$z$, respectively. This multiplied signal is passed to a low-pass filter (LPF). For Butterworth filters, which digital filters approximate, the gain and phase shift are as shown in Equation \ref{th_gain}, \cite{horowitz1989art}

\begin{equation}
    \label{th_gain}
    \begin{split}
        G_n(\omega) = \frac{1}{\Big(1+(\omega \tau)^2\Big)^{n/2}},\quad \Theta_n(\omega) = -n \tan^{-1}(\omega \tau)
    \end{split}
\end{equation} 

Here, $n$ refers to the number of filter stages or the slope of the filter in the frequency domain and $\tau$ is the time constant of the filter, which is defined as $\tau = 1/(2 \pi f_c)$, where $f_c$ is the critical frequency (the frequency at which the filter attenuates a signal by $-3$ dB). In our experiments, the filter roll-off is 24 dB/oct or $n = 4$. 

The LIA output can be derived by Taylor expanding around $V_0$ and $z_0$, resulting in Equations \ref{th_eq_didv_output} and \ref{th_eq_didz_output}. Here, $\Delta \omega = |\omega_V - \omega_z|$, and  $\Delta \phi = |\phi_V - \phi_z|$.  The complete calculation can be found in the Appendix (Section \ref{app_lia}).

    \begin{widetext}
        \begin{equation}
        \label{th_eq_didv_output}
        \begin{split}
            \text{Output}_{\text{d}I/\text{d}V} \approx& \frac{I(V_{0},z_{0})}{2}\Big(\rho(eV_0,V_0,0)\hat{v} + 2G_4(\omega_V)\sin(\omega_{V}t - \Theta_4(\omega_V))\\
            &-2 \kappa (V_0) \hat{z} G_4(\Delta \omega) \cos(\Delta \omega t + \Delta \phi - \Theta_4(\Delta \omega))\\
            &+ \kappa (V_0) \hat{z}\rho(eV_0,V_0,0)\hat{v} \Big[ G_4(\omega_V - \Delta \omega)\sin((\omega_V-\Delta \omega) t + \phi_V - \Delta \phi - \Theta_4(\omega_V-\Delta \omega))\\ 
            &- G_4(\omega_z)\sin(\omega_zt + \phi_z - \Theta_4(\omega_z)) \Big] \Big)\\
        \end{split}
        \end{equation}  

        \begin{equation}
        \label{th_eq_didz_output}
        \begin{split}
            \text{Output}_{\text{d}I/\text{d}z} \approx& \frac{I(V_{0},z_{0})}{2}\Big(-2 \kappa (V_0) \hat{z} + 2G_4(\omega_z)\sin(\omega_{z}t - \Theta_4(\omega_z))\\
            &+ \rho(eV_0,V_0,0)\hat{v}G_4(\Delta \omega)\sin(\Delta \omega t + \Delta \phi)\\
            &+ \kappa (V_0) \hat{z}\rho(eV_0,V_0,0)\hat{v} \Big[ G_4(\omega_z - \Delta \omega)\sin((\omega_z-\Delta \omega) t + \phi_z - \Delta \phi - \Theta_4(\omega_z-\Delta \omega))\\
            &- G_4(\omega_V)\sin(\omega_Vt + \phi_V - \Theta_4(\omega_V)) \Big] \Big)\\
        \end{split}
        \end{equation}  
    \end{widetext}

A key assumption in the derivation of Equations \ref{th_eq_didv_output} and \ref{th_eq_didz_output} is that the oscillation amplitudes must be small. For d$I$/d$V$ this means that $\hat{v}<<V_0$, while for d$I$/d$z$ it is necessary that $\kappa \hat{z} << 1$.

We end up with four frequencies which might fall within the pass band of the LPF: $\omega = 0$,  $\omega = \Delta \omega$, $\omega = \omega_V - \Delta \omega$ and $\omega = \omega_z - \Delta \omega$, where $\Delta \omega = |\omega_V - \omega_z|$. To get an accurate output from the LIA every AC component should fall in the stop band of the LPF, so we would correctly obtain:

\begin{equation}
    \label{th_eq_ideal_outputs}
    \begin{split}
        \text{Output}_{\text{d}I/\text{d}V} \propto& I(V_{0},z_{0})\rho(eV_0,V_0,0),\\ 
        \text{Output}_{\text{d}I/\text{d}z} \propto& I(V_{0},z_{0})\kappa (V_0)
    \end{split}
\end{equation}

The general rule for all LPFs is that any signal for which Equation \ref{th_eq_art_lia} holds, is heavily attenuated. So, only components in the input signal which are within $1/(2\pi \tau)$ of the reference frequency will enter the output of the LIA unscathed.

\begin{equation}
\label{th_eq_art_lia}
\begin{split}
    \omega > \frac{1}{\tau}
\end{split}
\end{equation}

This means Equation \ref{th_eq_criterium} must be obeyed when selecting the frequencies at which the voltage and tip piezo are modulated, where $f_c=1/(2\pi\tau)$ is known as the critical angle and $\Delta f = |f_z - f_V|$. In our experiments we set $\tau = 1$ ms and so $f_c \approx 159$ Hz.

\begin{equation}
\label{th_eq_criterium}
\begin{split}
    &\Big[f_V\,\wedge\,f_z\,\wedge\,\Delta f\,\wedge\,(f_{V} - \Delta f)\,\wedge\,(f_{z} - \Delta f)\Big]> f_c
\end{split}
\end{equation}

\section{Experimental}
\label{experimental}
The STM system of choice is an Omicron ultra-high vacuum low-temperature STM. During the experiments, the base pressure in the STM main chamber is around $10^{-11}$ mbar. All measurements are carried out at a temperature of 77 K. We employ a guarded I-V converter to achieve low noise levels. The LIAs of choice are two Stanford Research Systems SR830s. These are digital lock-ins using digital Butterworth LPFs, which offer an intensely sharp transition from the passband to the stopband by allowing up to four filter stages to be used.

For the modulation signals, we follow three crucial criteria: (i) the amplitudes must be as small as possible, (ii) the frequencies must be well-above the cut-off frequency of the feedback loop of the STM electronics (> 1 kHz) during measurements where the feedback is engaged and (iii) the frequency difference has to be large enough to prevent the LIAs from interfering.

The samples we have used are flame-annealed Au(111) films ($200$ nm) on mica (Phasis, Geneva, Switzerland). We choose this substrate because of its ease-of-use and the presence of surface phenomena that will allow us to properly test and verify the dual modulation technique.

\begin{figure*}
    \includegraphics[width=\textwidth]{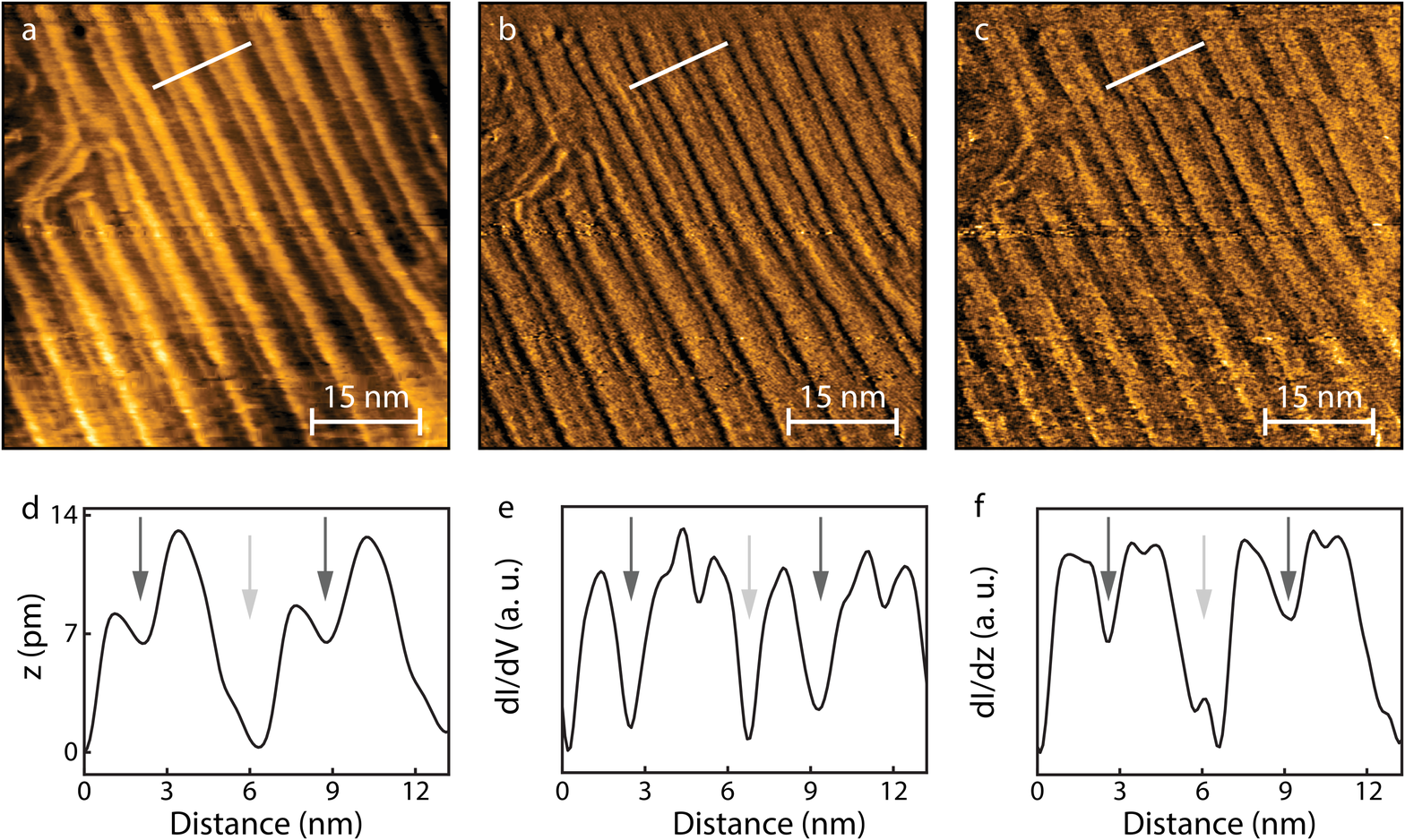}
    \caption{(a) Topography showing the herringbone reconstruction of the Au(111) surface. The image was captured using a bias voltage of $-50$ mV and a tunneling current setpoint of $400$ pA. (b) d$I$/d$V$ map of the herringbone measured using $\hat{v}=20$ mV and $f_{V}=2247$ Hz. (c) d$I$/d$z$ map of the herringbone for which $\hat{z}=0.27$ \si{\angstrom} and $f_{z} = 2929$ Hz. The topography and both spectroscopic maps were measured simultaneously. (d)-(f) Graphs showing lines profiles taken at the white lines. The line profiles are averaged over a 4 nm perpendicular window. The dark grey arrows indicate $hcp$ stacked regions and the light grey arrows mark $fcc$ stacked regions. (d) The line profile from the topography. The topographical protrusions in-between the $hcp$ and $fcc$ stacked regions are known as bridge regions. The height of these protrusions is roughly $10$ pm. (e) The line profile from the d$I$/d$V$ map. It is somewhat similar to the topographical line profile. (f) The line profile from the d$I$/d$z$ map. It shows a clear difference in the apparent tunnel barrier between the \textit{hcp} and \textit{fcc} regions.}
    \label{fig_maps} 
\end{figure*}

\section{Results and Discussion}

In Figure \ref{fig_maps}(a)-(c) the topography (a), d$I$/d$V$ map (b), and d$I$/d$z$ map (c) of a Au(111) surface are shown. The topography image and two maps are recorded simultaneously at a sample bias of $-50$ mV and a tunneling current set point of $400$ pA. To obtain the spectroscopic maps the bias was modulated using a $2247$ Hz signal with an amplitude of $20$ mV and the z-piezo was modulated using a $2929$ Hz signal with a $0.27$ \si{\angstrom} amplitude. Both frequencies are chosen according to the criteria stated in Section \ref{experimental}. Each image shows the presence of the well-known herringbone reconstruction of Au(111). The herringbone reconstruction has a ($22\,\times\,\sqrt{3}$) unit cell containing 23 atoms in the topmost layer and 22 atoms in the second layer. The resulting compression causes the stacking sequence to switch between $fcc$ and $hcp$, separated by bridge regions. \cite{woll1989determination}

In Figure \ref{fig_maps}(d)-(f), we show profiles taken across two herringbone ridges in the topography (d), d$I$/d$V$ (e), and d$I$/d$z$ (f) map. The d$I$/d$V$ map and topography are somewhat similar. This is expected, since the tunneling current at low $V$ results from an integration of the LDOS over a small energy window, and so the topography in a constant current image will contain information of the LDOS itself. The d$I$/d$z$ profile, which provides information on the local tunnel barrier, shows the tunnel barrier is higher on the $hcp$ regions than on the $fcc$ regions and even higher on the bridge regions. This shows that the stacking positions of the Au atoms affect the local tunnel barrier, matching results from other reports. \cite{aoki2014mapping}

\begin{figure}
    \includegraphics[width=0.5\textwidth]{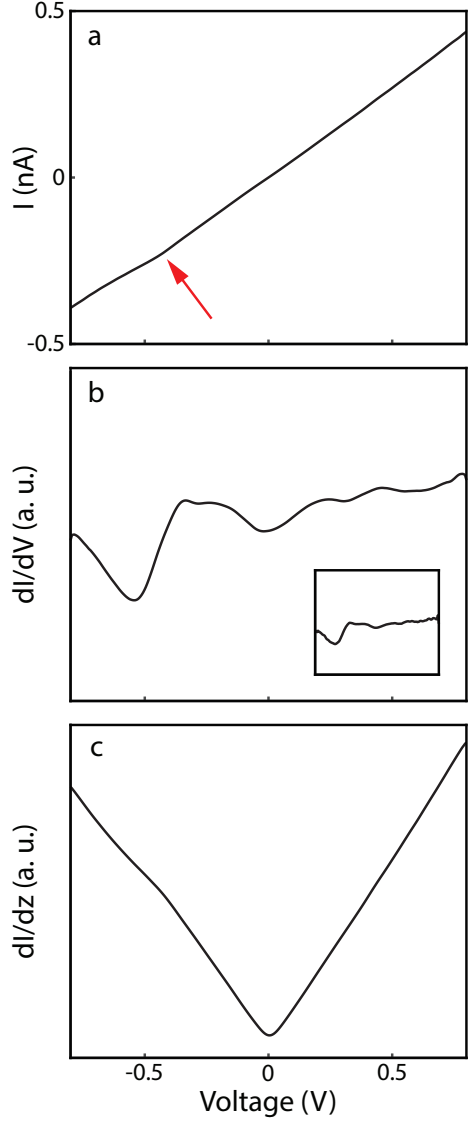}
    \caption{The averaged results of open-loop 20x20 grid $I$($V$) taken between $-0.8$ V to $0.8$ V. (a) The average current gives a linear curve, as is expected for a metal. The red arrow indicates a kink in the $I$($V$) curve around $-0.45$ V, which corresponds to the Au(111) surface state. (b) The average d$I$/d$V$, which was captured using $\hat{v}=20$ mV and $f_{V}=2117$ Hz. The surface state that was only faintly visible in the $I$($V$) curve translates to a step change in the d$I$/d$V$. The inset in Panel (b) shows the numerical derivative of (a) for comparison. (c) The average d$I$/d$z$ for which we used $\hat{z}=0.27$ \si{\angstrom} and $f_{z} = 2643$ Hz. The d$I$/d$z$ signal is directly proportional to the current, which shows that the inverse decay length is bias independent for Au(111). The polarity flip at $0$ V, and as such also the inversion of the kink around $-0.45$ V, is caused by the fact that we measure the modulus ($R$) which is always positive.}
    \label{fig_curves}
\end{figure}

In Figure \ref{fig_curves}(a)-(c) the result of an $I$($V$) grid scan is displayed. In panel (a) the average open-loop $I$($V$) curve is shown, whereas in panels (b) and (c) the average d$I$/d$V$ and d$I$/d$z$ signals as a function of sample bias are shown. The $I$($V$) curve shows a small kink, indicated by a red arrow, around a sample bias of $-0.45$ V. This kink corresponds to the location of the Au(111) 2D surface state \cite{kevan1987high}. In the d$I$/d$V$ curve this surface state shows up as a sudden increase in the LDOS and, as such, the d$I$/d$V$ signal. To emphasize this, the inset shows the numerical derivative of panel (a) for comparison. The d$I$/d$z$ signal, shown in panel (c), follows the shape of the $I$($V$) curve, including the kink caused by the surface state, with the exception that the polarity of the signal flips at $0$ V. This polarity flip is due to the fact that we measure the absolute value of the LIA output ($R$). As such, the inverse decay length, which can be extracted by dividing the d$I$/d$z$ by the tunneling current $I$, is found to be nearly independent of the sample bias.

\begin{figure*}
    \includegraphics[width=\textwidth]{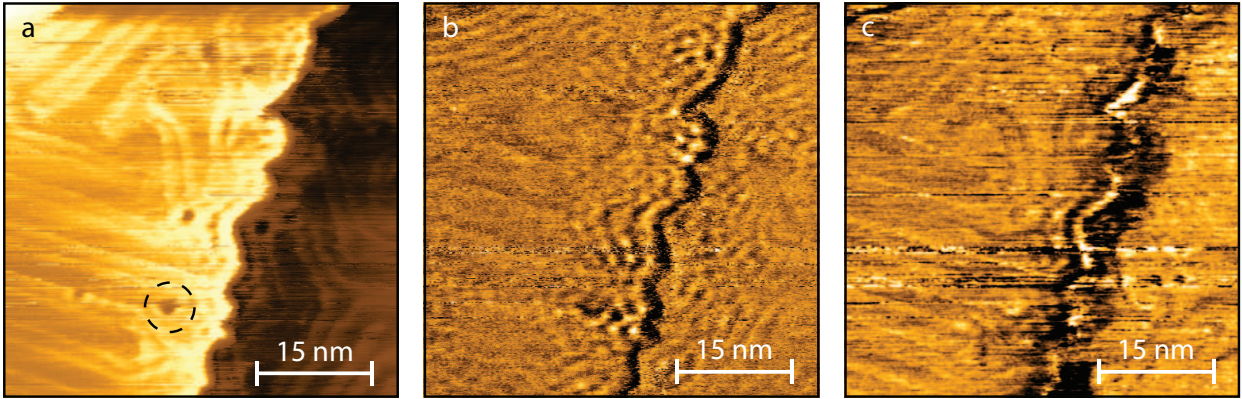}
    \caption{(a) Topography scan of a step edge, taken at $-100$ mV at a setpoint of $400$ pA. The black circle indicates one of the vacancy islands, which are one atomic layer deep. Despite the vacancy islands, the herringbone construction is preserved. (b) d$I$/d$V$ map of a step edge made using $f_V = 2117$ Hz and $\hat{v} = 20$ mV. Friedel oscillations are present around the vacancy islands, near the step edge in the middle of the image, and near another step edge in the top-left corner of the image. (c) d$I$/d$z$ map of a step edge recorded with $f_z = 2643$ Hz and $\hat{z} = 0.27$ \si{\angstrom}. The herringbone reconstruction is, although less so compared to Figure \ref{fig_maps}, visible in both spectroscopic maps. The topography and both spectroscopic maps were recorded simultaneously.}
    \label{fig_friedel}
\end{figure*}

Figure \ref{fig_friedel}(a)-(c) shows the topography (a), d$I$/d$V$ map (b), and d$I$/d$z$ map (c) of a step edge. The scan is made at a bias voltage of $-100$ mV and a tunneling current setpoint of $400$ pA. The d$I$/d$V$ map is recorded using a modulation frequency of $2117$ Hz and an amplitude of $20$ mV, while the d$I$/d$z$ map is recorded with a modulation frequency of $2643$ Hz and an amplitude of $0.27$ \si{\angstrom}. The frequencies are slightly different from experiment to experiment to optimize the signal-to-noise ratio.

In the vicinity of the step edge, several vacancy islands have formed, one of which is marked in (a) with a black circle. The herringbone reconstruction is clearly visible in the topography, albeit distorted by the step edge and vacancy islands, while it is also faintly visible in the d$I$/d$V$ and d$I$/d$z$ maps.

The d$I$/d$V$ map reveals Friedel oscillations on the upper terrace, lower terrace and especially around the vacancy islands. \cite{hasegawa1993direct} Additionally, on the top left, more Friedel oscillations are visible due to another step edge. The length of the Friedel oscillations is roughly $1.8$ nm, which is consistent with other studies. \cite{sotthewes2016method}

\section{Conclusion}

We have simultaneously measured d$I$/d$V$ and d$I$/d$z$ using an STM and two LIAs, both during topographic imaging and during open-loop experiments. We have shown from a theoretical standpoint that the two spectroscopic modes can be measured simultaneously without any interference between them as long as the modulation frequencies of $z$ and $V$ are chosen with care. This was supported by experimental results, as key electronic features such as the inverse decay length and changes in the LDOS were obtained from measured spectra. As such, this dual modulation technique allows for the acquisition of a wealth of spectroscopic information, and their relation to topographic features, in a relatively short amount of time.

\section{Acknowledgements}

This work is part of the research program 'X-tools - A thin-film physics toolbox for XUV optics', funded by the Netherlands Organisation for Scientific Research NWO, with co-funding by ASML, Carl Zeiss SMT, and Malvern Panalytical (NWO ref. 741.018.301). The authors acknowledge the support of the Industrial Focus Group XUV Optics and the Physics of Interfaces and Nanomaterials Group at the MESA+ Institute for Nanotechnology at the University of Twente.

\section{Data Availability}

The data that support the findings of this study are available from the corresponding author (v.j.s.oldenkotte@utwente.nl) upon reasonable request.

\section{References}

\providecommand{\noopsort}[1]{}\providecommand{\singleletter}[1]{#1}%

\newpage

\section{Appendix}

        \renewcommand{\theequation}{A\arabic{equation}}   
        \setcounter{equation}{0}
        
        \subsection{Lock-in Output Derivation \label{app_lia}}
        The local tunneling current $I$ can be approximated within the Tersoff-Hamann model \cite{tersoff1983theory} as a function of the tip distance $z$ and bias voltage $V$ as,
        
        \begin{equation}
            I(V,z) = C\int_0^{eV} \rho(\epsilon,V,z)\, d\epsilon.
        \label{app_eq_TersoffHamann}
        \end{equation}
        
        Using Leibniz's rule, this gives us the differentials $\partial I/\partial V$ following,
        
        \begin{equation}
            \frac{\partial I}{\partial V}(V,z) = e C \rho(eV,V,z)
            + C\int_0^{eV} \frac{\partial \rho}{\partial V}(\epsilon,V,z)\, d\epsilon ,
        \label{app_eq_didv1}
        \end{equation}
        
        and $\partial I/\partial z$ following,
        
        \begin{equation}
            \frac{\partial I}{\partial z}(V,z) = C \int_0^{eV} \frac{\partial \rho}{\partial z}(\epsilon,V,z)\, d\epsilon ,
        \label{app_eq_didz1}
        \end{equation}
        
         All energies are measured with respect to $E_F$, which we set to zero for simplicity. Furthermore, electronic states are indicated using the band index and the Bloch momentum $k_{||}$. The local density of states $\rho(\epsilon,V,z)$ of a crystalline surface, in the presence of the bias potential $V$ at the tip position $z$, is given by, 
         
        \begin{equation}     
        \begin{split}
             \rho(\epsilon,V,z) =& \sum_n\int \delta(\epsilon-\epsilon_{n\mathbf{k}_\parallel}(V)) \big|\psi_{n\mathbf{k}_\parallel}(V,z)\big|^2\, d\mathbf{k}_\parallel \\
             =& \sum_n\int_{C_{n\epsilon}} \frac{\big|\psi_{n\mathbf{k}_\parallel}(V,z)\big|^2}{\big|\nabla_{\mathbf{k}_\parallel}\epsilon_{n\mathbf{k}_\parallel}(V)\big|} \, d l
        \label{app_eq_rho}
        \end{split}
        \end{equation}
         
         In this expression, the 2D-integration variable is changed from $\mathbf{\kappa}_{||}$ to $(\epsilon,l)$, where $l$ runs over the constant energy contours $C_{n \epsilon}$, and $1/| \nabla \epsilon |$ is the corresponding Jacobian for this transformation. The remaining bias dependence of the energies and one-electron states is caused by the bias potential in the barrier, which will penetrate and be partially screened by the crystal surface. As a first order approximation, we may include only the change in the tails of the electronic wave functions in the barrier, caused by the bias voltage, and consider the one-electron states and energies as undistorted otherwise. For $z>>0$, these wave functions will show an exponential decay in $z$, with an amplitude that is proportional to its value at the surface,
        
        Here we only consider the lowest order Fourier component in $\mathbf{r}_{||}$, as this will dominate in the barrier. The decay parameter $\kappa_{n \mathbf{k}_{||}}(V)$ satisfies the parallel-momentum and energy-matching relation,
        
        \begin{equation}
            \frac{\hbar^2}{2m} \Big( k^2_\parallel- \kappa^2_{n\mathbf{k}_\parallel}(V) \Big) + U(V)  = \epsilon_{n\mathbf{k}_\parallel}.
        \end{equation}
        
        The barrier $U(V)$ can be estimated as the average $\overline{\phi}$ of the work functions of the sample $\phi_s$ and the tip material $\phi_t$, modified by the bias voltage $V$,
        
        \begin{equation}
            U(V)= \frac12\big(\Phi_s + \Phi_t + eV\big) = \bar\Phi + \frac12eV.
        \end{equation} 
        
        This gives,
        
        \begin{equation}
        \begin{split}
            \kappa_{n\mathbf{k}_\parallel}(V) =& \sqrt{\frac{2m}{\hbar^2} \big(\bar\Phi + \frac12eV  - \epsilon_{n\mathbf{k}_\parallel}\big) + k^2_\parallel}\\
            \Rightarrow &
            \frac{\partial \kappa_{n\mathbf{k}_\parallel}}{\partial V}(V) = 
            \frac{me}{2\hbar^2}\frac1{\kappa_{n\mathbf{k}_\parallel}(V)}
        \end{split}
        \end{equation}
        
        To first order, we thus get an approximation on the constant-energy surface,
        
        \begin{equation}
        \begin{split}
            \kappa_{n\mathbf{k}_\parallel}(V) =& \kappa(\epsilon,k_\parallel)+ \frac{m}{2\hbar^2}\frac1{\kappa(\epsilon,k_\parallel)}(eV-\epsilon), \text{ Where:}\\
            \kappa(\epsilon,k_\parallel) =& 
            \sqrt{\frac{2m}{\hbar^2} \big(\bar\Phi - \frac12 \epsilon\big) + k^2_\parallel}
        \end{split}
        \end{equation}
        
        While it suffices to include only the zeroth order term in factors $\kappa_{n \mathbf{k}_{||}}(V)$, we will need to include the first order term in the exponential, as these lead to scaling factors. Using these relations, we get,
        
        \begin{eqnarray}
            \rho(\epsilon,V,z) &=& 
             \sum_n\int_{C_{n\epsilon}} \big|\psi_{n\mathbf{k}_\parallel}\big|^2 \frac{e^{-2\kappa_{n\mathbf{k}_\parallel}(V) z}}{\big|\nabla_{\mathbf{k}_\parallel}\epsilon_{n\mathbf{k}_\parallel}\big|} \, d l
             \nonumber \\
             &\approx&
            e^{-\frac{m z (eV-\epsilon)}{\hbar^2 \bar\kappa_0(\epsilon)} }
            \underbrace{\sum_n\int_{C_{n\epsilon}} \big|\psi_{n\mathbf{k}_\parallel}\big|^2 \frac{e^{-2\kappa(\epsilon,k_\parallel) z}}{\big|\nabla_{\mathbf{k}_\parallel}\epsilon_{n\mathbf{k}_\parallel}\big|} \, d l}_{\rho(z,\frac1e\epsilon,\epsilon)}
            \label{app_eq_simplify_rho}
        \end{eqnarray}
        
        with $\overline{\kappa}_0(\epsilon)$ an effective decay constant. The advantage of this approximation is that we only need to calculate the density of states once, and not for each bias value.
        
        We may approximate the derivatives of the local density of states in a similar fashion,
        
        \begin{equation}
        \begin{split}
            \frac{\partial \rho}{\partial z}(\epsilon,V,z) =& -2 \sum_n\int_{C_{n\epsilon}} \big|\psi_{n\mathbf{k}_\parallel}\big|^2 \frac{\kappa_{n\mathbf{k}_\parallel}(V) e^{-2\kappa_{n\mathbf{k}_\parallel}(V) z}}{\big|\nabla_{\mathbf{k}_\parallel}\epsilon_{n\mathbf{k}_\parallel}\big|} \, d l\\
            \approx& - 2 \bar\kappa_1(V,\epsilon) \rho(\epsilon,V,z)
        \label{app_eq_drhodz}
        \end{split}
        \end{equation}
        
        in which $\overline{\kappa}_1(V,\epsilon)$ is another effective decay parameter at a particular energy and bias, and
        
        \begin{eqnarray}
            \frac{\partial \rho}{\partial V}(\epsilon,V,z) 
            &=& 
            - \frac{me z}{\hbar^2} \sum_n\int_{C_{n\epsilon}} \big|\psi_{n\mathbf{k}_\parallel}\big|^2 \frac{e^{-2\kappa_{n\mathbf{k}_\parallel}(V) z}}
            {\kappa_{n\mathbf{k}_\parallel}(V) \big|\nabla_{\mathbf{k}_\parallel}\epsilon_{n\mathbf{k}_\parallel}\big|} \, d l
            \nonumber \\&\approx&
            - \frac{me z}{\hbar^2 \bar\kappa_2(V,\epsilon)} \rho(\epsilon,V,z)
        \label{app_eq_drhodv}
        \end{eqnarray}
        
        in which $\overline{\kappa}_2(V,\epsilon)$ is again an effective decay parameter at a particular energy and bias. For $|eV|\ll\bar\Phi$, $\bar\kappa_1(V,\epsilon) \approx \bar\kappa_2(V,\epsilon)$, and we can consider them measures of the same effective decay length.
        
        Now, assuming $\partial z / \partial V \approx 0$ , combining Equations \ref{app_eq_didv1}, \ref{app_eq_didz1}, \ref{app_eq_drhodz}, and \ref{app_eq_drhodv} gives us the differentials d$I$/d$V$ following,
        
        \begin{equation}
        \begin{split}
            \frac{dI}{dV}(V,z) \approx& \rho(eV,V,z) - \frac{mez}{\hbar^2 \kappa(V)}I(V,z)
        \end{split}
        \end{equation}
        
        and d$I$/d$z$ following,
        
        \begin{equation}
             \frac{dI}{dz}(V,z) \approx -2 \kappa(V) I(V,z)               
        \label{app_eq_didz2}
        \end{equation}        
        
        Where $\kappa(V)$ is a weighted average over the energy range $0$ to $eV$ of $\bar\kappa_1(V,\epsilon)$. The second term in the d$I$/d$V$ relates to the changes in the shape of the electronic wave functions in the vacuum gap caused by the electric field. \cite{feenstra1987tunneling} However, this term is generally much smaller than the LDOS. Additionally, under the assumption that $\kappa(V)$ does not vary significantly over $V$, the LDOS at height $z$ may be approximated further as shown below,
        
        \begin{equation}
        \begin{split}
            \frac{dI}{dV}(V,z) \approx& \rho(eV,V,z) \approx \rho(eV,V,0)I(V,z)
        \label{app_eq_didv2}
        \end{split}
        \end{equation}        
        
        In our experiment, the derivatives of the tunneling current are obtained by applying a small sinusoidal modulation is to the bias voltage and $z$-piezo voltage. The result is two oscillating functions: one for the bias voltage and one for the piezo height.
        
        \begin{equation}
            \begin{split}
                V = V_{0} + \hat{v}\sin(\omega_{V}t + \phi_{V}), \quad \frac{\hat{v}}{V_0} << 1,\\ 
                z = z_{0} + \hat{z}\sin(\omega_{z}t + \phi_{z}), \quad \kappa \hat{z} << 1 \\
                \label{app_eq_modulations}
            \end{split}
        \end{equation}
        
        Due to the small amplitude of both modulations, the expansion shown in Equation \ref{app_approx_small_mod} can be applied to both variables:

        \begin{equation}
            \begin{split}
                f(x) =& f(x_0 + \hat{x}sin(\omega x t + \phi_x))\\ 
                \approx& f(x_0) + \hat{x}sin(\omega x t + \phi_x)\frac{d f}{d x}\bigg\rvert_{x=x_{0}}
                \label{app_approx_small_mod}
            \end{split}
        \end{equation}
        
        By combining Equations \ref{app_eq_modulations} and \ref{app_approx_small_mod}, the time-dependent current is obtained.
        
        \begin{equation}
            \begin{split}
                I(t) \approx& I(V_{0},z_{0}) \Big( 1 + \hat{v}\sin(\omega_{V} t + \phi_V)\frac{dI}{dV}\bigg\rvert_{V=V_0,z=z_0}\\
                &+  \hat{z}\sin(\omega_{z} t + \phi_z)\frac{dI}{dz}\bigg\rvert_{V=V_0,z=z_0}\\
                &+ \hat{v}\hat{z}\sin(\omega_{V} t + \phi_V)\sin(\omega_{z} t + \phi_z)\frac{d^2I}{dVdz}\bigg\rvert_{V=V_0,z=z_0}\Big)
                \label{app_eq_it}
            \end{split}
        \end{equation}

        Next, Equations \ref{app_eq_didv2} and \ref{app_eq_didz2} can be entered into Equation \ref{app_eq_it}.
        
        \begin{equation}
            \begin{split}
                I(t) \approx& I(V_{0},z_{0}) \Big( 1 + \rho(eV_0,V_0,0)\hat{v}\sin(\omega_{V} t + \phi_V)\\
                &-  2\kappa (V_0)\hat{z}\sin(\omega_{z} t + \phi_z)\\
                &- 2\kappa (V_0) \hat{z}\rho(eV_0,V_0,0)\hat{v}\sin(\omega_{V} t + \phi_V)\sin(\omega_{z} t + \phi_z)\Big)
                \label{app_eq_it_filled_in}
            \end{split}
        \end{equation}        
        
        The lock-in amplifier retrieves the signal around a certain frequency $\omega_{ref}$, by multiplying the incoming signal with the reference. Information in the input around $\omega_{ref}$ will translate to a DC signal and as such, applying a LPF will select this signal. To calculate the output of a lock-in amplifier, often the limit of $t>>\tau$ is used to obtain a time-averaged integral\cite{mandelis1994signal}: $\frac{1}{T}\int_0^T f_{ref}(t)f_{signal}(t)dt$. However, since $f \propto 1/\tau$ this limit is optimal around DC, so we cannot use this approximation to analyze interference which, as we will show, consists of AC components.
        
        We will only show all steps for the derivation of the d$I$/d$V$ output, since the calculation for the d$I$/d$z$ output is analogous. We introduce $\Delta \omega = |\omega_V - \omega_z|$ and $\Delta \phi = |\phi_V - \phi_z|$ to improve readability. Using $2\sin(a)\sin(b) = \cos(a-b) - \cos(a+b)$ and discarding all $\cos(a+b)$ terms (since they will certainly be filtered away by the LPF) gives us Equation \ref{app_eq_multiplied}.
        
        \begin{widetext}
            \begin{equation}
                \label{app_eq_multiplied}
                \begin{split}
                    \sin(\omega_{V}t+\phi_V)I(t) \approx& \frac{I(V_{0},z_{0})}{2}\Big( 2\sin(\omega_{V}t) +\rho(eV_0,V_0,0)\hat{v} -2 \kappa (V_0) \hat{z}\cos(\Delta \omega t + \Delta \phi)\\
                    &- 2\kappa (V_0) \hat{z}\rho(eV_0,V_0,0)\hat{v} \sin(\omega_V t + \phi_V) \Big[ \cos(\Delta \omega t + \Delta \phi)\\
                    &- \cos((\omega_z + \omega_V)t + \phi_V + \phi_z) \Big] \Big)\\
                    =& \frac{I(V_{0},z_{0})}{2}\Big( 2\sin(\omega_{V}t) +\rho(eV_0,V_0,0)\hat{v} -2 \kappa (V_0) \hat{z}\cos(\Delta \omega t + \Delta \phi)\\
                    &- 2\kappa (V_0) \hat{z}\rho(eV_0,V_0,0)\hat{v} \sin(\omega_V t + \phi_V) \Big[ \sin(\Delta \omega t + \Delta \phi - \frac{\pi}{2})\\
                    &- \sin((\omega_z + \omega_V)t + \phi_V + \phi_z - \frac{\pi}{2}) \Big] \Big)\\
                    \approx& \frac{I(V_{0},z_{0})}{2}\Big( 2\sin(\omega_{V}t) +\rho(eV_0,V_0,0)\hat{v} -2 \kappa (V_0) \hat{z}\cos(\Delta \omega t + \Delta \phi)\\
                    &+ \kappa (V_0) \hat{z}\rho(eV_0,V_0,0)\hat{v} \Big[ \sin((\omega_V-\Delta \omega) t + \phi_V - \Delta \phi) - \sin(\omega_zt + \phi_z )\Big] \Big)\\
                \end{split}
            \end{equation}
        \end{widetext}
        
        Because digital lock-in amplifiers can use more than two filter stages, without losing a significant amount of signal or resolution, they can afford to use simplistic filter architectures with flat pass and stop bands while maintaining a sharp transition. So, the digital LPF can be described by a multiplication in the frequency domain of a signal with a Butterworth transfer function as shown in Equation \ref{app_butterworth}.
        
        \begin{equation}
        \label{app_butterworth}
        \begin{split}
             H(\omega) =& \frac{1}{1 + j\omega \tau}
        \end{split}
        \end{equation}
        
        The gain and phase shift that result from this type of filter are well-known. \cite{horowitz1989art} But, in case the reader is not too familiar with electronics, we provide a derivation using a dummy sine function $f(t) = \sin(kt + \phi)$. We start by taking the Fourier transform,
        
        \begin{equation}
        \label{app_fourier_sine}
        \begin{split}
             f(t) =& \sin(kt + \phi) = \frac{1}{2j}\Big( e^{j(kt + \phi)} - e^{-j(kt + \phi)} \Big)\\
             F(\omega) =& \frac{1}{2j}\Big( e^{j\phi}\delta(\omega - k) - e^{-j\phi}\delta(\omega + k) \Big)
        \end{split}
        \end{equation}
        
        We now multiply Equations \ref{app_butterworth} and \ref{app_fourier_sine} to obtain the filtered output function and then perform an inverse Fourier transform to calculate the convolution in the time domain.

        \begin{widetext}        
            \begin{equation}
            \label{app_output}
            \begin{split}
                 H(\omega)F(\omega) =& \frac{1}{2j}\frac{1}{1+j\omega \tau}\Big( e^{j\phi}\delta(\omega - k) - e^{-j\phi}\delta(\omega + k) \Big)\\
                 h(t)*f(t) =& \frac{1}{2j}\int^{\infty}_{-\infty} \Big( \frac{e^{j\phi}\delta(\omega - k)}{1+j\omega \tau} -  \frac{e^{-j\phi}\delta(\omega + k)}{1+j\omega \tau} \Big) e^{j \omega t}d\omega\\
                 =& \frac{1}{2j}\Big( \frac{e^{j(kt+\phi)}}{1+jk\tau} - \frac{e^{-j(kt+\phi)}}{1-jk\tau}\Big)\\
                 =& \frac{1}{2j}\Big( \frac{e^{j(kt+\phi)}(1-jk\tau) - e^{-j(kt+\phi)}(1+jk\tau)}{1 + (k \tau)^2}\Big)\\
            \end{split}
            \end{equation}
        \end{widetext}
        
        Now we use Euler's formula: $z = x + iy \, \rightarrow \, z = r e^{i \theta}$, where: $r = \sqrt{x^2 + y^2}$ and $\theta = tan^{-1}(y/x)$.

        \begin{equation}
            \label{app_output_2}
            \begin{split}        
                 h(t)*f(t)=& \frac{1}{2j}\frac{e^{j(kt+\phi)}e^{j\tan^{-1}(-k \tau)} - e^{-j(kt+\phi)}e^{j\tan^{-1}(k \tau)}}{\sqrt{1+(k \tau)^2}}\\
                 =& \frac{1}{2j}\frac{e^{j(kt+\phi - \tan^{-1}(k \tau))} - e^{-j(kt+\phi - \tan^{-1}(k \tau))}}{\sqrt{1+(k \tau)^2}}\\
                 =& \frac{1}{\sqrt{1 + (k \tau)^2}} \sin(kt + \phi - \tan^{-1}(k \tau))\\
            \end{split}
        \end{equation}
        
        This is the result for a single LPF stage. For multiple stages Equation \ref{app_output_2} has to be iterated $n$ times. This leads to Equation \ref{app_output_n_stage} for an $n$-stage LPF, where $\Theta_n(\omega)$ is the frequency dependent phase shift and $G_n(\omega)$ is the gain.
        
        \begin{equation}
        \label{app_output_n_stage}
        \begin{split}        
             H_1(\omega)^nF(\omega) =& H_n(\omega)F(\omega) \rightarrow h_n(t) * f(t) \\
            =&G_{n}(\omega)\sin(kt + \phi - \Theta_n(\omega)),\\
             \text{Where: }\,G_n(\omega) =& \frac{1}{\Big(1+(\omega \tau)^2\Big)^{n/2}},\,\Theta_n(\omega) = n \tan^{-1}(\omega \tau)
        \end{split}
        \end{equation}
        
        Combining Equation \ref{app_eq_multiplied} and Equation \ref{app_output_n_stage}, we now obtain the output for the d$I$/d$V$ shown in Equation \ref{app_eq_didv_output}. The output for the d$I$/d$z$ can be calculated analogously to obtain Equation \ref{app_eq_didz_output}
        
        From Equation \ref{app_output_n_stage} one can see that if $k \tau > 1$ for a certain term, that term will be heavily attenuated. To measure the d$I$/d$V$ signal properly we need $\text{Output}_{\text{d}I/\text{d}V}\approx \rho(eV_0,V_0,0)\hat{v}$. Similarly, for the d$I$/d$z$ we need $\text{Output}_{\text{d}I/\text{d}z}\approx-2\kappa (V_0) \hat{z}$. This requires three conditions to be met: each of which boils down to a certain frequency being $f>f_c$, where $f_c$ is known as the critical frequency: $f_c = 1/(2\pi\tau)$. In our experiments $\tau$ was 1 ms and so this critical frequency was: $f_c = 1/(2\pi 10^{-3}) = 159$ Hz. 
        
        The first condition ($\omega_z > 2\pi f_c\,\wedge\,\omega_V > 2\pi f_c$) is realized automatically when using frequencies above the cut-off frequency of the STM electronics. However, if one were to measure with the feedback loop off, for instance while taking an $I$($V$) curve, this might give trouble if the operator chooses frequencies which are too low. The second condition ($\Delta \omega > 2\pi f_c $) will lead to interference when the modulations have similar frequencies, but can be satisfied by a proper choice of the modulation frequencies. The third condition requires that the lowest frequency minus the difference of the frequencies must also fall in the stopband: $\omega_z - \Delta \omega > 2\pi f_c\,\wedge\,\omega_V - \Delta \omega \tau > 2\pi f_c$. Here, we obtained the second part of the condition by anagolously performing the calculation for $\text{Output}_{\text{d}I/\text{d}V}$. Of course, as long as the above condition holds for the lowest frequency, it will also be true for the highest frequency. The way to interpret this is that if $f_z=2f_V$ then $f_z - \Delta f = 0$ and there will be interference. This condition does not increase restrictions on which frequencies can be picked, as it is fulfilled automatically if one takes into account that applying an external modulation to a non-linear electronic system can create harmonics and so the modulation frequencies should not be picked as multiples of each other.

        \begin{widetext}
            \begin{equation}
            \label{app_eq_didv_output}
            \begin{split}
                \text{Output}_{\text{d}I/\text{d}V} =& h_{4}(t) \Big( \sin(\omega_V t + \phi_V)I(t) \Big)\\ 
                \approx& \frac{I(V_{0},z_{0})}{2}\Big(\rho(eV_0,V_0,0)\hat{v} + 2G_4(\omega_V)\sin(\omega_{V}t - \Theta_4(\omega_V))\\
                &-2 \kappa (V_0) \hat{z} G_4(\Delta \omega) \cos(\Delta \omega t + \Delta \phi - \Theta_4(\Delta \omega))\\
                    &+ \kappa (V_0) \hat{z}\rho(eV_0,V_0,0)\hat{v} \Big[ G_4(\omega_V - \Delta \omega)\sin((\omega_V-\Delta \omega) t + \phi_V - \Delta \phi - \Theta_4(\omega_V-\Delta \omega))\\
                    &- G_4(\omega_z)\sin(\omega_zt + \phi_z - \Theta_4(\omega_z)) \Big] \Big)\\
            \end{split}
            \end{equation}  
  
            \begin{equation}
            \label{app_eq_didz_output}
            \begin{split}
                \text{Output}_{\text{d}I/\text{d}z} =& h_{4}(t) \Big( \sin(\omega_z t + \phi_z)I(t) \Big)\\ 
                \approx& \frac{I(V_{0},z_{0})}{2}\Big(-2 \kappa (V_0) \hat{z} + 2G_4(\omega_z)\sin(\omega_{z}t - \Theta_4(\omega_z))\\
                &+ \rho(eV_0,V_0,0)\hat{v}G_4(\Delta \omega)\sin(\Delta \omega t + \Delta \phi)\\
                    &+ \kappa (V_0) \hat{z}\rho(eV_0,V_0,0)\hat{v} \Big[ G_4(\omega_z - \Delta \omega)\sin((\omega_z-\Delta \omega) t + \phi_z - \Delta \phi - \Theta_4(\omega_z-\Delta \omega))\\
                    &- G_4(\omega_V)\sin(\omega_Vt + \phi_V - \Theta_4(\omega_V)) \Big] \Big)\\
            \end{split}
            \end{equation}  
        \end{widetext}

\end{document}